\documentclass[journal]{IEEEtran}

\usepackage{amsmath}
\usepackage{graphicx}
\usepackage{mathtools}
\usepackage{amsfonts}
\usepackage{pifont}
\usepackage{amssymb}
\usepackage{epstopdf}
\usepackage{color}
\usepackage[utf8]{inputenc}
\usepackage{setspace}
\usepackage{ragged2e}
\usepackage{epsfig}
\usepackage{tabu}
\usepackage{textgreek}
\usepackage{enumitem}
\usepackage{lipsum}
\usepackage{mathtools}
\usepackage{cuted}

\makeatletter
\newcommand{\vast}{\bBigg@{1.2}}
\newcommand{\Vast}{\bBigg@{2.3}}
\newcommand{\vastl}{\bBigg@{4}}
\newcommand{\Vastl}{\bBigg@{5}}
\newcolumntype{M}[1]{>{\centering\arraybackslash}m{#1}}
\newcolumntype{N}{@{}m{0pt}@{}}

\makeatother
\ifCLASSINFOpdf
 
\else
  
\fi

\begin{document}

\title{Unified Composite Distribution and Its Applications to Double Shadowed $\alpha-\kappa-\mu$ Fading Channels}

\author{Hussien Al-Hmood,~\IEEEmembership{Member,~IEEE,} 
         and H. S. Al-Raweshidy,~\IEEEmembership{Senior Member,~IEEE}
\thanks{Manuscript received November 00, 2020.}
\thanks{The authors are with the Department of Electronic and Computer Engineering, College of Engineering, Design and Physical Sciences, Brunel University London, UK, e-mail: Hussien.Al-Hmood@brunel.ac.uk, Hamed.Al-Raweshidy@brunel.ac.uk.}}

\markboth{Submitted for Publication,~Vol.~00, ~NO.~00, November 2020}%
{Author 1 \MakeLowercase{\textit{et al.}}: Bare Demo of IEEEtran.cls for Journals}

\maketitle

\begin{abstract}
In this paper, we propose a mixture Gamma shadowed (MGS) distribution as a unified composite model via representing the shadowing by an inverse Nakagami-$m$. Accordingly, the exact expression and the asymptotic behaviour at high average signal-to-noise ratio (SNR) regime of the fundamental statistics of a MGS distribution are derived first. These statistics are then applied to analyze the performance of the wireless communications systems over double shadowed $\alpha-\kappa-\mu$ fading via providing the outage probability (OP), the average bit error probability (ABEP), the average channel capacity (ACC), and the effective capacity (EC). The average area under the receiver operating characteristics curve (AUC) of energy detection (ED) is also analyzed. The numerical and simulation results are presented to verify the validation of our analysis. 
\end{abstract} 
\IEEEpeerreviewmaketitle
\section{Introduction}
\IEEEPARstart{T}{he} wireless communications channel may undergo the effect of the multipath and shadowing fading simultaneously [1]. Accordingly, many works have been recently dedicated to analyse the performance of the generalized composite models, such as, $\kappa-\mu$, $\eta-\mu$, and $\alpha-\mu$ distributions that are used to model the line-of-sight (LoS), the non-LoS (NLoS), and the non-linear wireless communication mediums, respectively [2], [3]. These generalized conditions can provide close results to the practical measurements and  approximately comprise all the classical fading distributions, i.e., Rayleigh, Nakagami-$m$, Nakagami-$n$, Nakagami-$q$, one-sided Gaussian, Weibull, and Gamma. Hence, the probability density function (PDF), the cumulative distribution function (CDF), and the moment generating function (MGF) of the composite $\eta-\mu$/Gamma fading models were derived in [4]. The authors in [5] assumed that both $\kappa-\mu$ and $\eta-\mu$ fading conditions are shadowed by an inverse Gamma distribution. The fundamental statistics of the $\kappa-\mu$ shadowed fading in which the shadowing effect is represented by a Nakagami-$m$ distribution were given in [6] with applications to the outage probability (OP) and the average bit error probability (ABEP) of wireless communications systems. The composite of the $\alpha-\kappa-\mu$/Gamma distribution was analysed in [7]. The non-linear scenario of the Fisher-Snedecor $\mathcal{F}$ distribution [8], namely, $\alpha-\mathcal{F}$, was investigated in [9]. The PDF, the CDF, and the MGF of the $\alpha-\eta-\mathcal{F}$ and the $\alpha-\kappa-\mathcal{F}$ composite fading conditions were derived in [10]. These models were then unified in a single distribution which is named $\alpha-\eta-\kappa-\mathcal{F}$ composite fading [11]. 
\par Although different statistical properties have been reported in [4]-[11], they are derived in mathematically intractable expressions. This is because they are expressed in terms of either the hypergeometric or the modified Bessel functions. Therefore, the mathematical intricacy of the performance metrics of the wireless communications systems is high and this would to unclear insights into the system behaviour with the fading parameters. Moreover, this complexity is increased when the generalised multipath fading channels undergo to double shadowing impacts [12]. 
\par To overcome the aforementioned challenges, a mixture Gamma (MG) distribution has been widely used to approximate with high accuracy most of the composite generalised/Gamma fading conditions [13]-[16]. For instance, the average area under the receiver operating characteristics (AUC) curve of the energy detection (ED) based spectrum sensing was derived in [14]. The average channel capacity (ACC) of $\alpha-\eta-\lambda-\mu$/Gamma fading and the effective capacity (EC) of $\alpha-\eta-\mu$/Gamma fading were analyzed in [15] and [16], respectively, using a MG distribution.  
\par Based on the above observations and motivated by the merits of a MG distribution, we propose a mixture Gamma shadowed (MGS) as a unified composite model where the shadowing is represented by an inverse Nakagami-$m$. To this end, the basic statistics of this distribution are mathematically simple and tractable. Consequently, a clear insight into the effects of the fading parameters on the performance of the wireless communications systems can be deduced when the channel is subjected to double shadowing impacts.     
\par Our main contributions are summarized as follows: 
\begin{itemize}
\item The exact and the asymptotic expressions at high average signal-to-noise ratio (SNR) values for novel unified mathematically tractable statistical characterizations of composite model that is based on MG and inverse Nakagami-$m$ distributions, namely, MGS distribution, are derived.  
\item The derived statistics are employed to analyse the performance of the double shadowed $\alpha-\kappa-\mu$ fading channel in which the first and the second shadowing impacts are respectively followed the Nakagami-$m$ and the inverse Nakagami-$m$ distributions. Accordingly, the equivalent parameters of a MG distribution of composite $\alpha-\kappa-\mu$/Nakagami-$m$ fading channels are provided.
\item Capitalizing on the above, unified closed-form expressions for the ABEP, the ACC, the EC, and the average AUC are obtained.
\end{itemize}   
\section{MG and Inverse Nakagami-$m$ Distributions}
The PDF of a MG distribution is given by [13, eq. (1)]
\label{eqn_1}
\setcounter{equation}{0}
\begin{align}
f(x)= \sum_{j=1}^{K} \sigma_j x^{\beta_j-1} e^{-\zeta_j x}
\end{align}
where $\sigma_j$, $\beta_j$, and $\zeta_j$ are the parameters of $j$th Gamma component and $K$ that stands for the number of terms is evaluated via using the mean square error (MSE) method between the exact PDF and its MG representation [13]. 
\par If $\xi$ is an inverse Nakagami-$m$ random variable (RV), its PDF    is expressed as [12, eq. (50)]
\label{eqn_2}
\setcounter{equation}{1}
\begin{align}
f_{\xi}(r) = \frac{(m_s-1)^{m_s}}{\Gamma(m_s)} r^{m_s-1} e^{-\frac{(m_s-1)}{r}}.  \quad m_s > 1
\end{align}
where $m_s$ refers to the shadowing severity index and $\Gamma(.)$ is the incomplete Gamma function [17, eq. (8.310.1)].
\section{Statistical Properties of a MGS Distribution}
\par Let $\gamma\sim \text{MGS}(\sigma_j, \beta_j, \zeta_j, m_s)$ for $j=1,\cdots, K$ is an MGS-distributed RV. Then, the PDF of $\gamma$ can be obtained by the product of MG and inverse Nakagami-$m$ RVs as [4, eq. (8)] 
\label{eqn_3}
\setcounter{equation}{2}
\begin{align}
f_{\gamma}(\gamma) = \int_0^\infty \frac{1}{z}f\bigg(\frac{\gamma}{z}\bigg) f_{\xi} (z) dz.
\end{align}
\par Substituting (1) and (2) into (3) and making use of [17, eq. (3.381.4)] with some mathematical manipulations, this yields 
\label{eqn_4}
\setcounter{equation}{3}
\begin{align}
f_{\gamma}(\gamma) = (m_s-1)^{m_s} \sum_{j=1}^{K}  \frac{\sigma_j (m_s)_{\beta_j} \gamma^{\beta_j-1}}{(m_s-1+\zeta_j \gamma)^{\beta_j+m_s}}.
\end{align}
where $(.)_n$ is the Pochhammer symbol.
\par  When $\zeta_j \rightarrow 0$ for all $j = 1,\cdots,K$, the asymptotic of the PDF, $f^{\infty}_{\gamma}(\gamma)$, can be expressed as
\label{eqn_5}
\setcounter{equation}{4}
\begin{align}
f^{\infty}_{\gamma}(\gamma) \simeq \sum_{j=1}^{K}  \frac{\sigma_j (m_s)_{\beta_j} \gamma^{\beta_j-1}}{(m_s-1)^{\beta_j}}.
\end{align}
\par Inserting (4) in $F_{\gamma}(\gamma)=\int_0^\gamma f_{\gamma}(\gamma) d\gamma$ and recalling [17, eq. (3.194.1)], the CDF of a MGS distribution can be derived as 
\label{eqn_6}
\setcounter{equation}{5}
\begin{align}
F_{\gamma}(\gamma) = &\sum_{j=1}^{K}  \frac{\sigma_j (m_s)_{\beta_j} {_2F_1}\big(\beta_j+m_s,\beta_j;\beta_j+1;-\frac{\zeta_j}{m_s-1}\gamma\big)}{ \beta_j ((m_s-1)\gamma^{-1})^{\beta_j}}.
\end{align}
where ${_2F_1}(.)$ is the Gauss hypergeometric function [17, eq. (9.14.1)].
\par Using the fact that ${_2F_1}(.,.;.;0) \simeq 1$ when $\zeta_j \rightarrow 0$ or plugging (5) in $F_{\gamma}(\gamma)=\int_0^\gamma f_{\gamma}(\gamma) d\gamma$, the asymptotic of the CDF, $F^{\infty}_{\gamma}(\gamma)$, can be evaluated as 
\label{eqn_7}
\setcounter{equation}{6}
\begin{align}
F^{\infty}_{\gamma}(\gamma) \simeq  \sum_{j=1}^{K}  \frac{\sigma_j (m_s)_{\beta_j} \gamma^{\beta_j}}{ \beta_j(m_s-1)^{\beta_j}}.
\end{align}
\par Using the Laplace transform and invoking [18, eq. (13.2.5)], the MGF of the MGS distribution can be obtained as  
\label{eqn_8}
\setcounter{equation}{7}
\begin{align}
\mathcal{M}_{\gamma}(s) &= \sum_{j=1}^{K}  \frac{\sigma_j (m_s)_{\beta_j} \Gamma(\beta_j) U\big(\beta_j;1-m_s;\frac{m_s-1}{\zeta_j}s\big)}{ \zeta^{\beta_j}_j} .
\end{align}
where $U(.)$ is the Tricomi confluent hypergeometric
function of the second kind defined in [17, eq. (9.211.4)].
\par The asymptotic of the MGF, $\mathcal{M}^{\infty}_{\gamma}(s)$, can be deduced after applying Laplace transform for (5) and invoking [17, eq. (8.310.1)]. Thus, this yields 
\label{eqn_9}
\setcounter{equation}{8}
\begin{align}
\mathcal{M}^{\infty}_{\gamma}(s) \simeq  \sum_{j=1}^{K}  \frac{\sigma_j (m_s)_{\beta_j} \Gamma(\beta_j)}{ [(m_s-1)s]^{\beta_j}}.
\end{align}
\par The $n$-th moment, $\boldsymbol{\mu}_n$, of the MGS distribution can be found by using (4) and [17, eq. (3.194.3)] as  
\label{eqn_10}
\setcounter{equation}{9}
\begin{align}
\boldsymbol{\mu}_n=\mathbb{E}\{\gamma^n\} = (m_s-1)^n \sum_{j=1}^{K}  \frac{\sigma_j (m_s)_{\beta_j}}{\zeta^{\beta_j+n}_j} B(\beta_j+n, m_s-n).
\end{align}
where $B(.)$ is the Beta function [17, eq. (8.380.1)].
\par It is worth interesting that the MGS distribution can used to model the $\kappa-\mu$/inverse Gamma [5, eq. (6)] with $\theta_j=\frac{e^{-\mu \kappa}\mu^{\mu+2(j-1)}\kappa^{j-1}}{\Gamma(\mu+j-1)\Gamma(j)\bar{\gamma}^{\mu+j-1}}$ which is used to compute $\sigma_j$, $\beta_j=\mu+j-1$, and $\zeta_j = \frac{\mu(1+\kappa)}{\bar{\gamma}}$. Additionally, the Fisher Senedcor $\mathcal{F}$ composite fading [8, eq. (6)] can be represented by (4)-(10) with $K = 1$, $\sigma_1 = \frac{m^m}{\Gamma(m) \bar{\gamma}^m}$, $\beta_1 = m$, and $\zeta_1 = \frac{m}{\bar{\gamma}}$.
\label{eqn_14}
\setcounter{equation}{13}
\begin{table*}[t]
\begin{align}
&f(x)= \frac{\alpha \mu m^m \kappa^\frac{1-\mu}{2} (1+\kappa)^\frac{1+\mu}{2}}{2 \Gamma(m) \exp(\mu \kappa) \bar{\gamma}^\frac{\alpha(1+\mu)}{4}} x^{\frac{\alpha(1+\mu)}{4}-1} \int_0^\infty z^{-\frac{\alpha(1+\mu)}{4}+m-1} e^{-\frac{(1+\kappa)\mu x^{\alpha/2}}{(\bar{\gamma} z)^{\alpha/2}}-m z}  I_{\mu-1} \bigg(2\mu \sqrt{\frac{\kappa(1+\kappa) x^{\alpha/2} }{(\bar{\gamma} z)^{\alpha/2}}}\bigg) dz.
\end{align}
\hrulefill
\end{table*}
\section{Double Shadowed $\alpha-\kappa-\mu$ Fading Channels}
The received signal envelope, $R$, over double shadowed $\alpha-\kappa-\mu$ fading channel can be expressed as
\label{eqn_11}
\setcounter{equation}{10}
\begin{align}
R^\alpha= \xi^2 \sum_{l=1}^{\mu}(X_l+ \vartheta p_l)^2+(Y_l+ \vartheta q_l)^2
\end{align}
where the parameters of (11) are defined as follows:
\begin{enumerate}[label=\roman*)] 
\item $\alpha>0$ denotes the non-linearity of the propagation
medium.  
\item $\mu$ is a real-valued extension related to the number of multipath clusters.
\item $\xi$ and $\vartheta$ represent the RVs which are responsible for introducing the shadowing impacts that are modelled by inverse Nakagami-$m$ and Nakagami-$m$ distributions, respectively, with $\mathbb{E}[\xi^2] = \mathbb{E}[\vartheta^2] =1$, where $\mathbb{E}[.]$ stands for the expectation operator. It is worth mentioning that (11) becomes equivalent to example 1 of the double shadowed $\kappa-\mu$ type I model [12, eq. (22)], when $\alpha = 2$. 
\item $X_l$ and $Y_l$ are mutually independent Gaussian random processes with mean $\mathbb{E}[X_l]$ and $\mathbb{E}[Y_l] = 0$ and variance $\mathbb{E}[X^2_l]=\mathbb{E}[Y^2_l] = \delta^2$.
\item $p_l$ and $q_l$ are the mean values of the in-phase and quadrature phase components of the multipath cluster $l$.
\end{enumerate}
\par From (11), one can note that the PDF of $R$ can be derived by averaging the PDF of the $\alpha-\kappa-\mu$ fading over the PDF of $\vartheta$ and $\xi$ RVs. However, the PDF of the $\alpha-\kappa-\mu$ fading is included the modified Bessel function of the first kind, $I_\upsilon(.)$ [17, eq. (8.445)] that would lead to mathematically intractable statistical properties (please refer to [12]). Therefore, to obtain simple closed-form statistics, the PDF of inducing the shadowing of the dominant component is approximated by using a MG distribution whereas the multiplicative shadowing is added by utilizing a MGS model. 
\par The PDF of the instantaneous SNR, $\gamma$, over double shadowed $\alpha-\kappa-\mu$ fading is expressed as [7, eq. (4)] 
\label{eqn_12}
\setcounter{equation}{11}
\begin{align}
f_\gamma(r)= \frac{\alpha \mu \kappa^\frac{1-\mu}{2} (1+\kappa)^\frac{1+\mu}{2}}{2 \exp(\mu \kappa) \bar{\gamma}^\frac{\alpha(1+\mu)}{4}} r^{\frac{\alpha(1+\mu)}{4}-1} e^{-\frac{(1+\kappa)\mu}{\bar{\gamma}^{\alpha/2}} r^{\alpha/2}} \nonumber\\
\times I_{\mu-1} \bigg(2\mu \sqrt{\frac{\kappa(1+\kappa)}{\bar{\gamma}^{\alpha/2}} r^{\alpha/2}}\bigg).
\end{align}
where $\kappa$ is the ratio between the total powers of the dominant components and scattered waves and $\bar{\gamma}$ is the average SNR.
\par The PDF of $\vartheta$ is given by [12, eq. (54)]
\label{eqn_13}
\setcounter{equation}{12}
\begin{align}
f_{\vartheta}(r) = \frac{m^m}{\Gamma(m)} r^{m-1} e^{-m r}. 
\end{align}
where $m$ is the shadowing severity index of the Nakagami-$m$.
\par Substituting (12) and (13) into (3), we have (14) shown at the top of this page. 
\par Using the substitution $y=\frac{(1+\kappa)\mu x^{\alpha/2}}{(\bar{\gamma} z)^{\alpha/2}}$ in (14), we obtain
\label{eqn_15}
\setcounter{equation}{14}
\begin{align}
f(x)= \Xi x^{m-1}\int_0^\infty e^{-y}  h(y)dy.
\end{align}
where $\Xi=\frac{m^m \mu^{\frac{2}{\alpha}m-\frac{\mu-1}{2}} \kappa^\frac{1-\mu}{2} (1+\kappa)^{\frac{2}{\alpha}m}}{\Gamma(m) \exp(\mu \kappa) \bar{\gamma}^m}$ and $h(y) = y^{\frac{\alpha(1+\mu)}{4}-\frac{2}{\alpha}m-1} e^{-\frac{m ((1+\kappa)\mu)^{\alpha/2}}{\bar{\gamma} y^{2/\alpha}} x}  I_{\mu-1} \big(2 \sqrt{\kappa \mu y}\big)$.
\par The integration in (15), $\Phi= \int_0^\infty e^{-y} h(y) dy$, can be approximated by using a Gaussian-Laguerre quadrature method, as $\Phi \approx \sum_{j=1}^{K} w_{j} h(y_j) $, where $w_{j}$ and $y_{j}$ are the weight factors and abscissas, respectively, given in [18]. Hence, (15) can be equivalently expressed by (1) with the following coefficients
 \label{eqn_16}
\setcounter{equation}{15}
\begin{align}
\beta_j = m,& \quad \zeta_j = \frac{m ((1+\kappa)\mu)^{2/\alpha}}{\bar{\gamma} y^{2/\alpha}_j}, \quad \sigma_j=\frac{\theta_j}{\sum_{l=1}^{K} \theta_l \Gamma(\beta_l) \zeta^{-\beta_l}_l} \nonumber\\
&\theta_j= \Xi w_j y^{\frac{\alpha(1+\mu)}{4}-\frac{2}{\alpha}m-1}_j I_{\mu-1} \big(2 \sqrt{\kappa \mu y_j}\big).
\end{align}
\par From (16), one can see that $\zeta_j \rightarrow 0$ when $\bar{\gamma} \rightarrow \infty$.  
\section{Performance Analysis using a MGS Model}
\subsection{Outage Probability}
The OP is defined as the probability of falling the values of the output SNR below a predefined threshold value $\varphi$. 
\par The OP, $P_o$, can be computed by [1, eq. (1.4)]
 \label{eqn_17}
\setcounter{equation}{16}
\begin{align}
P_o = F_\gamma(\varphi).
\end{align}
where $F_\gamma(.)$ is provided in (6).
\par The asymptotic of the OP, $P^{\infty}_o$, can be analysed by (7), i.e., $P^{\infty}_o = F^{\infty}_\gamma(\varphi)$. Furthermore, the $P^{\infty}_o$ may be closely represented as $P^{\infty}_o \simeq \bar{\gamma}^{-G_d}$ whereby $G_d$ denotes the diversity gain that demonstrates the increasing in the slope of the OP versus $\bar{\gamma}$. Hence, plugging $\sigma_j$ of (16) in (7), one can notice that the values of the $G_d$ of the MGS and double shadowed $\alpha-\kappa-\mu$ are proportional to $\beta_j$ and $m$, respectively.
\label{eqn_34}
\begin{table*}[t]
\setcounter{equation}{33}
\begin{align}
\bar{\mathcal{A}}  = 1-\sum_{l=0}^{u-1} \sum_{i=0}^l {{l+u-1} \choose {l-i}} \frac{2^{-(l+i+u)}}{ i!} (m_s-1)^i \sum_{j=1}^{K} \sigma_j (m_s)_{\beta_j} \Gamma(\beta_j+i) U\bigg(\beta_j+i;i-m_s+1;\frac{m_s-1}{2\zeta_j}\bigg).
\end{align} 
\hrulefill
\end{table*}
\subsection{Average Bit Error Probability}
The ABEP can be evaluated by [1, eq. (9.11)] 
 \label{eqn_18} 
 \setcounter{equation}{17}
\begin{align}
\bar{P}_e = \frac{1}{\pi} \int_0^{\frac{\pi}{2}} \mathcal{M}_\gamma \bigg(\frac{\rho}{\sin^2 \phi}\bigg) d\phi.
\end{align}
where $\rho=0.5$, $\rho=1$, and $\rho=0.715$ for coherent BFSK, BPSK, and BFSK with minimum correlation, respectively.
\par Substituting (8) into (18) and employing the identity [19, eq. (07.33.26.0004.01)], we have  
 \label{eqn_19} 
 \setcounter{equation}{18}
\begin{align}
\bar{P}_e = \frac{1}{\pi \Gamma(m_s)}\sum_{j=1}^{K}  \frac{\sigma_j}{ \zeta^{\beta_j}_j} \int_0^{\frac{\pi}{2}} G^{2,1}_{1,2}\bigg[ \frac{(m_s-1)\rho}{\zeta_j \sin^2 \phi} \bigg\vert
\begin{matrix}
  1-\beta_j\\
  0, m_s\\
\end{matrix} 
\bigg] d \phi.
\end{align}
\par By employing the definition of the Meijer's $G$-function [19, eq. (07.34.02.0001.01)] and $t=\sin^2 \phi$, (19) becomes
 \label{eqn_20} 
 \setcounter{equation}{19}
\begin{align}
\bar{P}_e &= \frac{1}{2 \pi \Gamma(m_s)} \sum_{j=1}^{K}  \frac{\sigma_j}{ \zeta^{\beta_j}_j} \int_0^1 \frac{1}{\sqrt{(1-t)t}}\frac{1}{2 \pi i}\int_{\mathcal{L}} \Gamma(r) \nonumber\\
& \Gamma(m_s+r) \Gamma(\beta_j-r) \bigg(\frac{(m_s-1)\rho}{t \zeta_j }\bigg)^{-r} dr dt.
\end{align}
where $i=\sqrt{-1}$ and $\mathcal{L}$ is the suitable contours in the $r$-plane from $\varrho-i\infty$ to $\varrho+i\infty$ with $\varrho$ is a constant value. 
\par Changing the order of the integrations of (20) and then using [17, eq. (3.191.3)] for the linear integral, we obtain
 \label{eqn_21} 
 \setcounter{equation}{20}
\begin{align}
\bar{P}_e = \frac{1}{2 \pi \Gamma(m_s)}&\sum_{j=1}^{K}  \frac{\sigma_j}{ \zeta^{\beta_j}_j}  \frac{1}{2 \pi i}\int_{\mathcal{L}} \Gamma(r) \Gamma(m_s+r) \Gamma(\beta_j-r) \nonumber\\
& B(r+0.5,0.5) \bigg(\frac{(m_s-1)\rho}{\zeta_j }\bigg)^{-r}dr.
\end{align}
\par Recalling the properties [17, eq. (8.384.1)/ eq. (8.338.2)] and using [19, eq. (07.34.02.0001.01)], (21) is expressed as
\label{eqn_22} 
 \setcounter{equation}{21}
\begin{align}
\bar{P}_e = \frac{1}{2 \sqrt{\pi} \Gamma(m_s)}\sum_{j=1}^{K}  \frac{\sigma_j G^{2,1}_{2,2}\bigg[ \frac{(m_s-1)\rho}{\zeta_j} \bigg\vert
\begin{matrix}
  1-\beta_j, 1\\
  0, m_s, 0.5\\
\end{matrix} 
\bigg]}{ \zeta^{\beta_j}_j} .
\end{align}
\par The asymptotic of the ABEP at high $\bar{\gamma}$ regime, $\bar{P}^{\infty}_e$, can be deduced after inserting (9) in (18) and using $t = \sin^2 \phi$ as 
 \label{eqn_23} 
 \setcounter{equation}{22}
\begin{align}
\bar{P}^{\infty}_e \simeq  \frac{1}{\pi} \sum_{j=1}^{K} \frac{\sigma_j (m_s)_{\beta_j} \Gamma(\beta_j)}{ [(m_s-1)\rho]^{\beta_j}} \int_0^{1} \frac{t^{\beta_j}}{\sqrt{(1-t)t}} dt.
\end{align}
\par The integration of (23) is recorded in [17, eq. (3.191.3)]. Thus, after some mathematical manipulations, this yields 
 \label{eqn_24} 
 \setcounter{equation}{23}
\begin{align}
\bar{P}^{\infty}_e \simeq \frac{1}{2\sqrt{\pi}}\sum_{j=1}^{K}  \frac{\sigma_j (m_s)_{\beta_j} \Gamma(\beta_j+0.5)}{ \beta_j [(m_s-1)\rho ]^{\beta_j}} .
\end{align}
\par It is evident from (24) that the $G_d$ of the MGS model is proportional to $\beta_j$, i.e., $\bar{P}^{\infty}_e \simeq \bar{\gamma}^{-\beta_j}$. Hence, the $G_d$ for the double shadowed $\alpha-\kappa-\mu$ fading is proportional to $m$.
\subsection{Average Channel Capacity}
According to Shannon's theorem, the normalized ACC, $\bar{C}$, can be determined by [8, eq. (26)]
\label{eqn_25}
\setcounter{equation}{24}
\begin{equation}
\bar{C} =\frac{1}{\ln(2)} \int_0^\infty \text{ln}(1+\gamma) f_\gamma(\gamma) d\gamma.
\end{equation}
\par Plugging (4) in (25) and making use the identity [19, eq. (01.04.26.0002.01)] to write the natural logarithm in terms of the Meijer's $G$-function, we obtain 
\label{eqn_26}
\setcounter{equation}{25}
\begin{align}
\bar{C}=&\frac{(m_s-1)^{m_s}}{\ln(2)} \sum_{j=1}^{K}  \sigma_j (m_s)_{\beta_j} \nonumber\\
&\times \int_0^\infty \frac{\gamma^{\beta_j-1}}{(m_s-1+\zeta_j \gamma)^{\beta_j+m_s}} G^{1,2}_{2,2}\bigg[ \gamma \bigg\vert
\begin{matrix}
  1,1\\
  1,0\\
\end{matrix} 
\bigg]  d\gamma.
\end{align}
\par Employing [17, eq. (7.811.5)], (26) can be expressed in exact closed-form as 
\label{eqn_27}
\setcounter{equation}{26}
\begin{align}
\bar{C}=\frac{1}{\ln(2)  \Gamma(m_s)} \sum_{j=1}^{K} \frac{ \sigma_j}{\zeta_j^{\beta_j} } G^{2,3}_{3,3}\bigg[ \frac{m_s-1}{\zeta_j} \bigg\vert
\begin{matrix}
 1-\beta_j, 1,1\\
  m_s,1,0\\
\end{matrix} 
\bigg].
\end{align}
 \par The asymptotic of the ACC for $\bar{\gamma}\rightarrow \infty$, $\bar{C}^{\infty}$, can be evaluated via [8, eq. (28)]
\label{eqn_28}
\setcounter{equation}{27}
\begin{align}
\bar{C}^{\infty} \simeq \frac{1}{\ln(2)}\frac{\partial}{\partial n} \mathbb{E}[\gamma^n] \bigg\vert_{n=0}.
\end{align} 
\par Substituting (10) into (28), computing the partial derivative, and setting $n=0$, $\bar{C}^{\infty}$ over MGS model is deduced as 
\label{eqn_29}
\setcounter{equation}{28}
\begin{align}
\bar{C}^{\infty} \simeq \sum_{j=1}^{K} \frac{ \sigma_j \Gamma(\beta_j) \bigg[\ln\big(\frac{m_s-1}{\zeta_j}\big)+\psi(\beta_j)-\psi(m_s)\bigg]}{\ln(2) \zeta_j^{\beta_j} }  .
\end{align}
where $\psi(.)$ is the digamma function [17, eq. (8.360.1)].
\subsection{Effective Capacity} 
In Shannon's theorem, the ACC has been measured under perfect quality of service (QoS). However, in the EC, the constraints of the QoS, such as, system delay, are taken into consideration [5]. 
\par The EC can be calculated by [15, eq. (4)]
\label{eqn_30}
\setcounter{equation}{29}
\begin{align}
\mathcal{R}=-\frac{1}{A}\text{log}_2 \bigg\{\int_0^\infty (1+\gamma)^{-A}f_\gamma(\gamma)d\gamma \bigg\}.
\end{align}
where $ A \triangleq \theta T B/\mathrm{ln}(2)$ with $\theta$,  $T$, and $B$ denote the delay exponent, the time and the bandwidth of the channel.
\par Inserting (4) in (30) and employing [17, eq. (3.197.1)], the expression of the EC over MGS distribution is yielded as 
\label{eqn_31}
\setcounter{equation}{30}
\begin{align}
\mathcal{R}&=-\frac{1}{A}\text{log}_2 \bigg\{\sum_{j=1}^{K}  \frac{\sigma_j (m_s)_{\beta_j} B(\beta_j,m_s+A)}{ (m_s-1)^{\beta_j}} \nonumber\\ 
&\times {_2F_1}\bigg(\beta_j+m_s,\beta_j;\beta_j+m_s+A;1-\frac{\zeta_j}{m_s-1}\bigg) \bigg\}.
\end{align} 
\par Plugging (5) in (30) and recalling [17, eq. (3.194.3)], the asymptotic of the EC at $\bar{\gamma} \rightarrow \infty$, $R^{\infty}$, is deduced as 
\label{eqn_32}
\setcounter{equation}{31}
\begin{align}
R^{\infty} \simeq -\frac{1}{A}\text{log}_2 \bigg\{\sum_{j=1}^{K}  \frac{\sigma_j (m_s)_{\beta_j}}{ (m_s-1)^{\beta_j}} B(\beta_j,A-\beta_j) \bigg\}.
\end{align} 
\subsection{Average AUC of Energy Detection} 
The AUC is a single figure of merit that is suggested as an alternative performance measure to the ROC curve [14].
\par The average AUC, $\bar{\mathcal{A}}$, can be computed by [20, eq. (20)/ eq. (21)]
  \label{eqn_33}
    \setcounter{equation}{32}
\begin{align}
\bar{\mathcal{A}} & =  1- \sum_{l=0}^{u-1} \sum_{i=0}^{l}  \frac{{{l+u-1} \choose {l-i}}}{2^{l+i+u}i!} \int_0^\infty  \gamma^i e^{-\frac{\gamma}{2}} f_\gamma(\gamma) d\gamma.
\end{align}
where ${a \choose b}=\frac{a!}{b! (a-b)!}$ is the binomial coefficient.
\par Plugging (6) in (33) and utilizing [18, eq. (13.2.5)], $\bar{\mathcal{A}}$ is yielded as in (34) given at the top of this page.
\par The average AUC at high $\bar{\gamma}$ value, $\bar{\mathcal{A}}^{\infty}$, can be evaluated after inserting (5) in (33) and invoking [17, eq. (3.381.4)] as
\label{eqn_35}
\setcounter{equation}{34} 
\begin{align}
&\bar{\mathcal{A}}^{\infty}  \simeq 1-\sum_{l=0}^{u-1} \sum_{i=0}^l  \sum_{j=1}^{K} \frac{\sigma_j (m_s)_{\beta_j} {{l+u-1} \choose {l-i}} \Gamma(\beta_j+i)}{2^{l+u-\beta_j} (m_s-1)^{\beta_j} i!}.
\end{align}
\section{Analytical and Simulation Results}
In this section, the numerical and the asymptotic results of the derived performance measures are presented for different scenarios. To achieve MSE $\leq 10^{-5}$, we have used $K = 15$.
\par Figs. 1-4 demonstrate the OP for $\varphi = 5$ dB, the ABEP for BPSK modulation, the comparison between the normalized ACC and the EC for $A = 3.5$, and the average complementary of AUC (CAUC), $1-\bar{\mathcal{A}}$, for $u=3$ versus $\bar{\gamma}$, respectively. Three cases for the shadowing index $m_s$, namely, heavy ($m_s = 1.5$), moderate ($m_s = 5.5$), and light ($m_s = 50$) are analyzed. 
\par From the provided results, it can be observed that the performance becomes better when $\alpha$, $\kappa$ and/or $\mu$ increase. This is because the high values of $\alpha$, $\kappa$, and $\mu$ indicate that the system tends to be linear, the total power of the dominant components is less than that of the scattered waves and large number of multipath clusters arrive at the receiver, respectively. Additionally, the increasing in $m$ and/or $m_s$ means the impacts of the first and/or the second shadowing on the received signal are low. However, $m$ has the high effect on the performance when $\bar{\gamma} \rightarrow \infty$ due to improving in the $G_d$. 
 \par In Fig. 3, one can note that the ACC is higher than the EC for the same scenario. This refers to the impact of the system delay on the EC whereas its ignored in the ACC. Moreover, the EC is related to the average AUC via $u$ [20]. This relationship explains how the low quality of the received signal by the unlicensed user would reduce the detectability of the ED.  
\par In all figures, a perfect agreement between the exact and the simulation results as well as the asymptotic at high $\bar{\gamma}$ can be noticed which proves the validation of our expressions.
 \label{Fig_1}
\begin{figure}[t]
\centering
  \includegraphics[width=3.2 in, height=2.03 in]{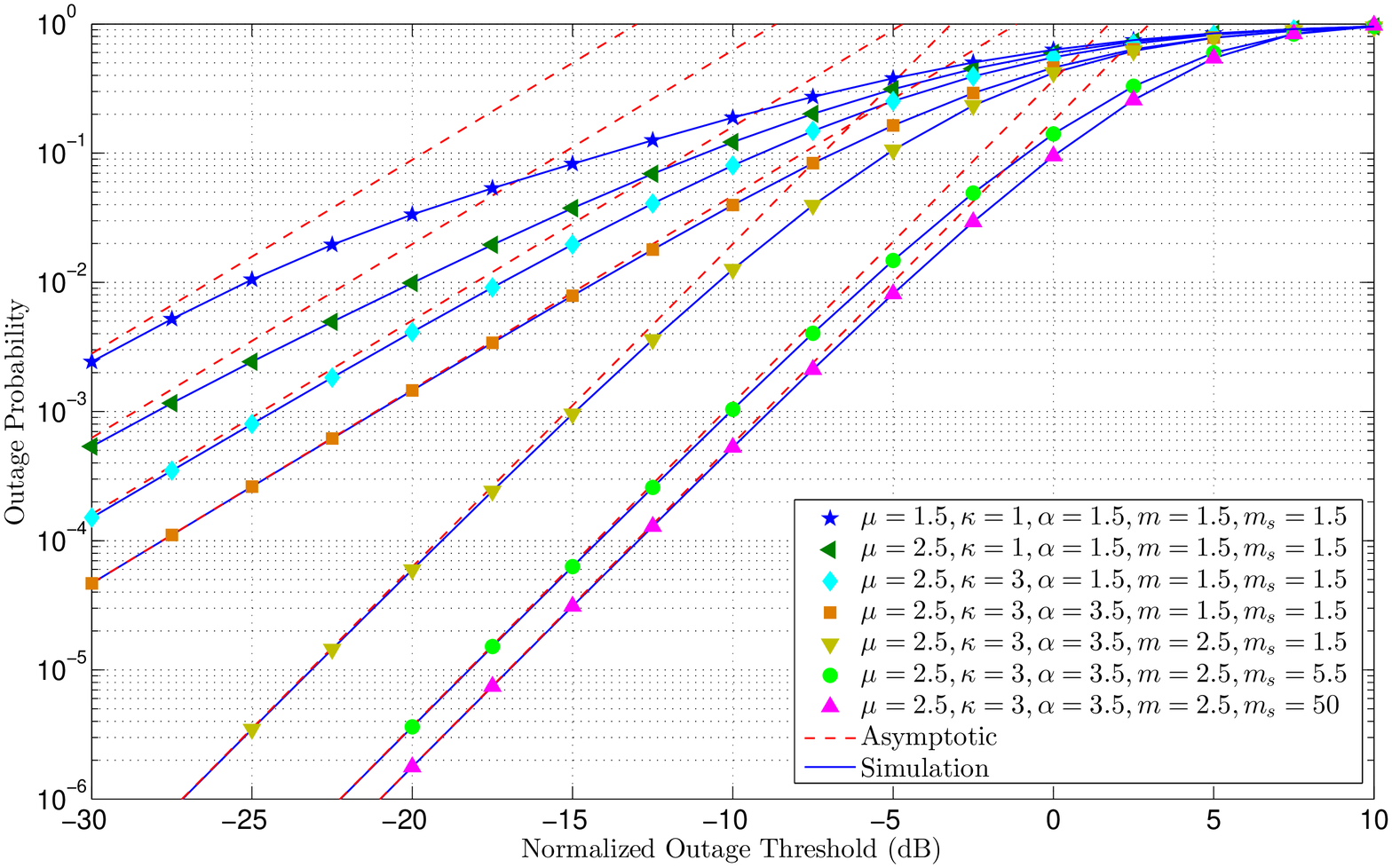}
\centering
\caption{OP versus normalized outage threshold for $\varphi = 5$ dB.}
\end{figure}
\label{Fig_2}
\begin{figure}[t]
\centering
  \includegraphics[width=3.2 in, height=2.03 in]{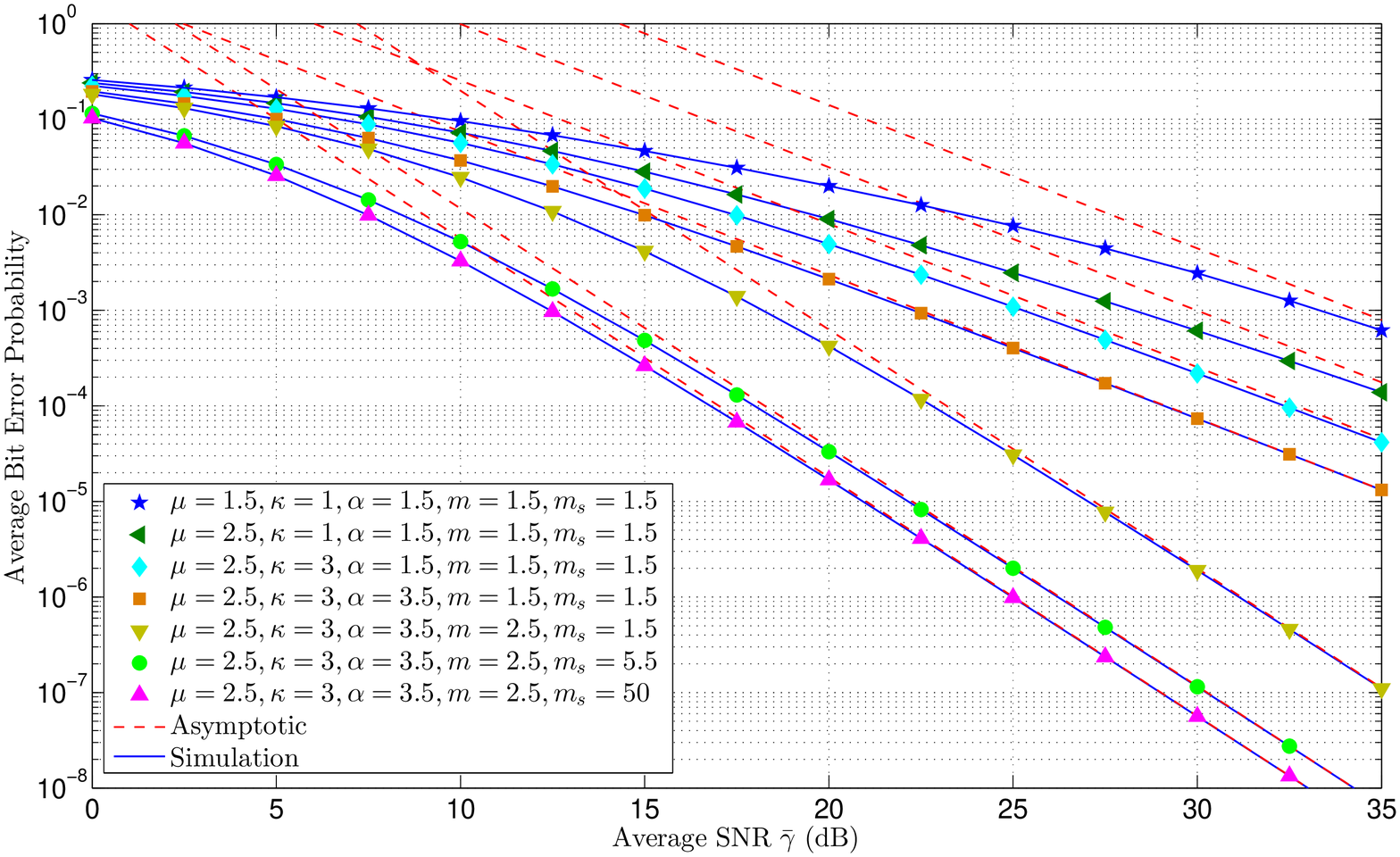}
\centering
\caption{ABEP for BPSK versus average SNR.}
\end{figure} 
\label{Fig_3}
\begin{figure}[t]
\centering
  \includegraphics[width=3.2 in, height=2.03 in]{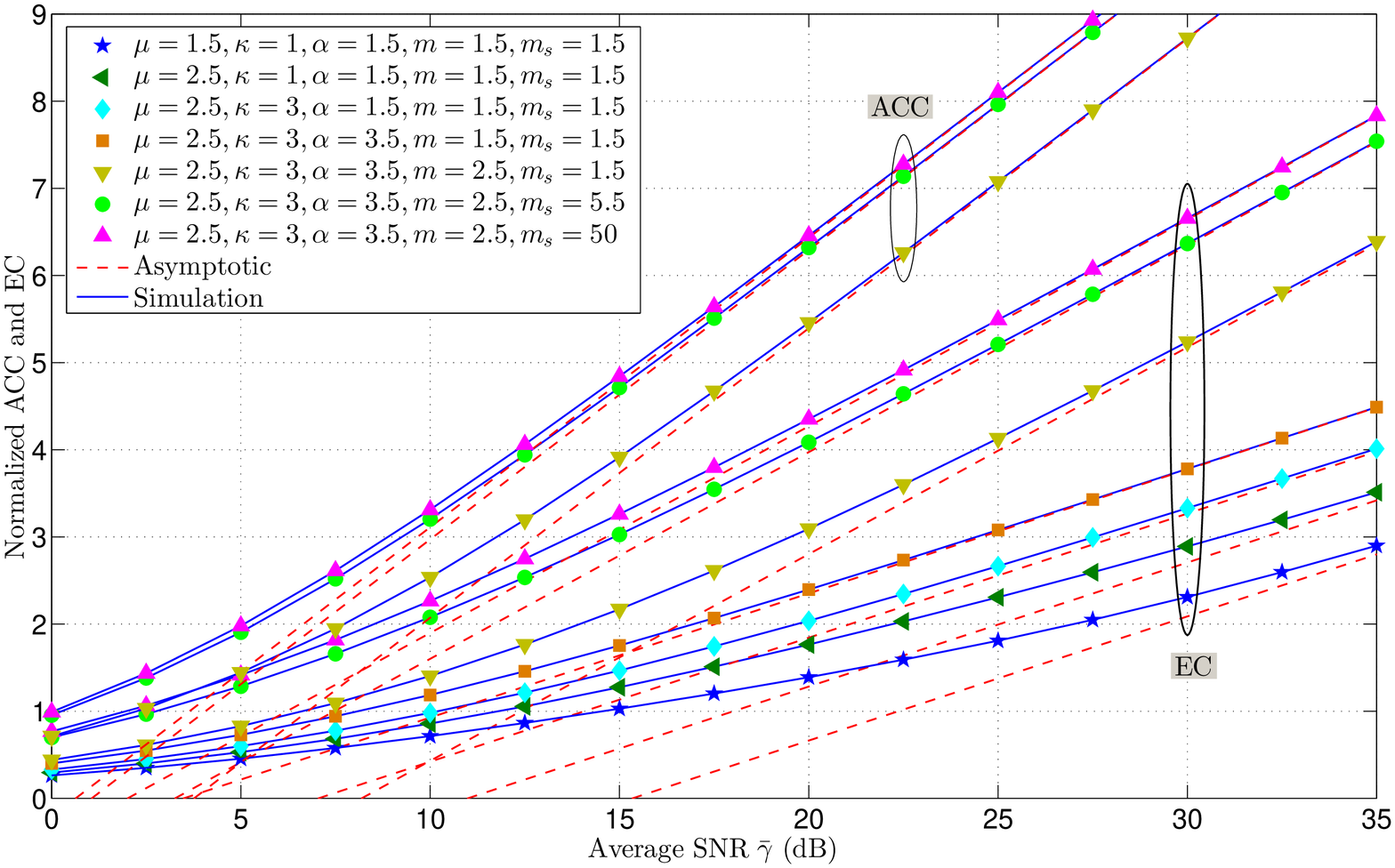} 
\caption{Normalized ACC and EC with $A = 3.5$ versus average SNR.}
\end{figure} 
\label{Fig_4}
\begin{figure}[t]
\centering
  \includegraphics[width=3.2 in, height=2.03 in]{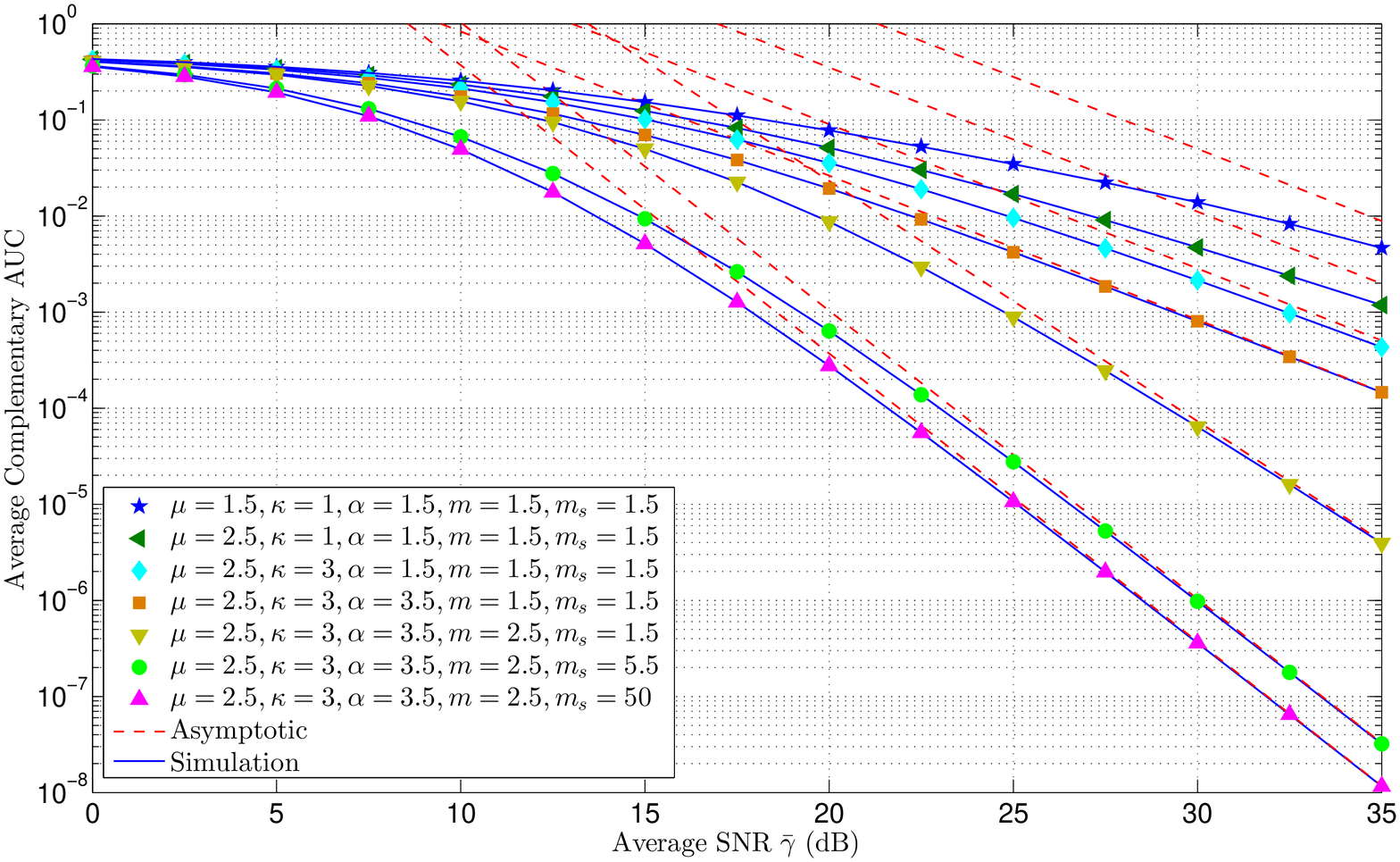} 
\centering
\caption{Complementary AUC for $u=3$ versus average SNR.}
\end{figure} 
\section{Conclusions}
In this paper, a MGS distribution was proposed as highly accurate approximate unified representation where the shadowing impact was assumed to be an inverse Nakagami-$m$ RV. This distribution was then applied to the double shadowed $\alpha-\kappa-\mu$ fading channel. The exact and the asymptotic expressions of the OP, the ABEP, the ACC, and the ER of the wireless communications systems and the average AUC of ED over a MGS model were derived. The results for different values of the fading parameters were presented. The expressions of this work can be employed for many fading conditions, such as, $\alpha-\kappa-\mathcal{F}$ fading with $m \rightarrow \infty$.

\ifCLASSOPTIONcaptionsoff
  \newpage
\fi

\end{document}